\documentstyle[prl,aps]{revtex} 
\input{psfig.sty} 
\tighten
\begin{document}
\draft
\twocolumn[
\hsize\textwidth\columnwidth\hsize\csname @twocolumnfalse\endcsname
\title{Boundary conditions for quasiclassical equations in the theory of
superconductivity} 
\author{ C. J. Lambert$^{*}$, R. Raimondi$^{*,a}$, V. Sweeney$^{*}$,
and A. F. Volkov$^{\dagger}$ }
\address{$^{*}$ School of Physics and Chemistry,
Lancaster University, Lancaster LA1 4YB, U.K.}
\address{$^{\dagger}$ Institute of Radioengineering and
Electronics of the Russian Academy of Sciences, Moscow}
\date{\today}
\maketitle
\begin{abstract}
In this paper we derive effective boundary conditions connecting the
quasiclassical Green's function through tunnel barriers in superconducting -
normal hybrid (S-N or S-S') structures in the dirty limit. Our work extends
previous treatments confined to the small transparency limit. This is
achieved by an expansion in the small parameter $r^{-1}=T/2(1-T)$ where T is
the transparency of the barrier. We calculate the next term in the $r^{-1}$
expansion for both the normal and the superconducting case. In both cases
this involves the solution of an integral equation, which we obtain
numerically. While in the normal case our treatment only leads to a
quantitative change in the barrier resistance $R_b$, in the superconductor
case, qualitative different boundary conditions are derived. To illustrate
the physical consequences of the modified boundary conditions, we calculate
the Josephson current and show that the next term in the $r^{-1}$ expansion
gives rise to the second harmonic in the current-phase relation.
\end{abstract}
] \narrowtext
\section{Introduction} \label{introduction}

In recent years there has been a great deal of work, both experimentally and
theoretically, on the transport properties of mesoscopic superconducting -
normal (N-S) hybrid structures. Various new effects have been observed
including zero-bias anomalies in the differential
conductance\cite{kastalsky91}, peculiar dependence on the magnetic field,
re-entrant temperature behaviour \cite{petrashov95}. All these phenomena are
caused by the interplay of Andreev scattering at N-S interfaces and phase
coherence in the mesoscopic region and are manifestations of the proximity
effect, whereby superconducting correlations are induced in the normal
region.

Among the various theoretical techniques used to deal with the above effects,
the quasiclassical method has revealed itself as one of the most powerful
approaches (see, for instance, \cite{larkin86} - \cite{volkov96}). This
technique enables the study of thermodynamical and kinetic properties of
superconductors (see \cite{larkin86} and references therein), whose
dimensions significantly exceed the Fermi wave length $\lambda_F =2\pi/k_F$
and where quantum size effects can be neglected. In the case of hybrid
structures the quasiclassical equations of motion must be supplemented by
appropriate boundary conditions in order to match the quasiclassical Green's
function ${\check g}$ at N-S interfaces. These have been derived by Zaitsev
\cite{zaitsev84} in a general form which is valid in both the clean and dirty
limits and take the form \cite{kuprianov88}

\begin{equation}
{\check a} \left( R -R {\check a}^2 + {T\over 4}
 ({\check s}_1 -{\check s}_2)^2 \right) =
{{T}\over 4} \left[ {\check s}_2, {\check s}_1 \right].
\label{1}
\end{equation}

Here ${\check a}$ and ${\check s}$ are the antisymmetric and symmetric parts
of the supermatrix Green's function ${\check g}$, i.e. 

\begin{equation}
{\check g}(\pm \mu )={\check s} \pm {\check a} ,
\label{2}
\end{equation}

where $\mu = cos(\theta ) = p_z / p_F$ and $\theta$ 
 is the angle between the velocity ${\bf
p}/m$ of an incident electron and the vector normal to the interface. The
reflection (R) and transmission (T) coefficients depend on $\mu$ and are
connected via the unitarity relation

\begin{equation}
R (\mu ) + T(\mu ) = 1.
\label{3}
\end{equation}

As shown in \cite{zaitsev84}, the antisymmetric part ${\check a}$ is
continuous at the interface, while the symmetric part ${\check s}$
experiences a jump determined by the commutator on the right hand side of
eq.(\ref{1})

\begin{equation}
\left[ {\check s}_2 , {\check s}_1 \right]=
{\check s}_2 {\check s}_1-{\check s}_1  {\check s}_2.
\label{4}
\end{equation}

The matrix elements of ${\check g}$ are the retarded (advanced) Green's
functions ${\hat g}^{R(A)}$ and the Keldysh Green's function  ${\hat g}$

\begin{equation}
{\check g} =
\left(
\begin{array} {c c}
{\hat g}^R & {\hat g} \\
0 & {\hat g}^A \\
\end{array}
\right).
\label{gmatrix}
\end{equation}

Following the usual convention, we denote two-by-two matrices in the Nambu
space by a "hat" (${\hat g}$) symbol and the four-by-four supermatrices by a
"check" (${\check g}$) symbol. The Keldysh Green's function ${\hat g}$
describes the kinetic effects and, by exploiting the normalization property,

$$
{\check g} {\check g} ={\check 1}
$$

is related to the matrix distribution function ${\hat f}$ via

\begin{equation}
{\hat g}={\hat g}^{R}{\hat f}-{\hat f}{\hat g}^{A}.
\label{5}
\end{equation}

In what follows we will adopt the convention introduced by Larkin and
Ovchinnikov, according to which the matrix ${\hat f}$ can be chosen to be
diagonal, with ${\hat f}=f_0 {\hat 1} + f_z {\hat \sigma}_z$.

The boundary conditions (\ref{1}) are valid in both the clean and dirty
limit. We recall that in the dirty limit ($l \ll \xi_{N,S}$,  $l$ is the mean
free path, $\xi_{N,S}$ are the coherence lengths in the N and S regions,
respectively) the angular dependence of the matrix ${\check g}$ can be taken
into account by keeping only the first two terms in an expansion in Legendre
polynomials $P_n (\mu )$. The first two terms in the expansion of ${\check
g}$ can be then related to each other. Far away from the interface this
relation can be written down as \cite{larkin86}

\begin{equation}
{\check a}_{\infty} =-l\mu {\check s}_{\infty} \partial_z {\check s}_{\infty}.
\label{6}
\end{equation}

\begin{figure}
{\psfig{figure=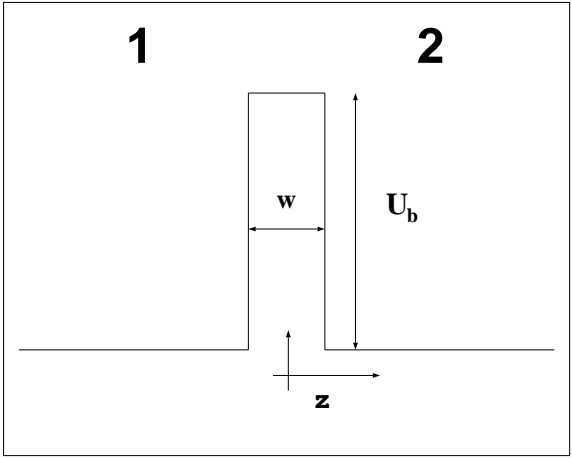,width=8cm}}
\caption{ Schematic picture of the structure studied in the text.}
\label{f1}
\end{figure}

The subscript $\infty$ means that $|z|\gg l$ (the interface is located at
$z=0$, see Fig.1). However, close to the interface, one should keep all the
terms in the expansion of ${\check g}$ because the coefficients $R$ and $T$
entering the boundary condition (\ref{1}) depend on $\mu$. In the dirty
limit, all higher terms ( $n\ge 2$) in the expansion of ${\check g}$ decay
exponentially with $z/l$, and it would be desirable to obtain a matching
condition at the interface which involves the asymptotic functions ${\check
a}_{\infty}$ and ${\check s}_{\infty}$ only. Such a problem was considered by
Kupriyanov and Lukichev \cite{kuprianov88}. They showed that the boundary
condition (\ref{1}) reduces then to

\begin{equation}
{\check a}_{2\infty}={\check a}_{1\infty}\equiv {\check a}_{\infty}
\label{7}
\end{equation}

\begin{equation}
{\check a}_{\infty}= {3\over 2} <\mu / r(\mu ) >
\left[ {\check s}_{2\infty} , {\check s}_{1\infty} \right].
\label{8}
\end{equation}

Here $r(\mu )=2R(\mu ) / T (\mu )$ and the angular brackets mean the angle
averaging

\begin{equation}
<\mu / r(\mu ) >= \int_0^1 ~~d\mu~\mu / r(\mu ). 
\label{9}
\end{equation}

The main aim of the present paper is to show that eq.(\ref{8}) is valid only
to lowest order in an expansion in the small parameter $r^{-1}$, i.e. in the
limit of a low transparency barrier, and consequently in the case of an
arbitrary barrier transmission, eq.(\ref{8}) can be used only for a
qualitative description. 

The layout of our paper is as follows. As a preliminary step, in the next
section, we consider the case of two normal regions separated by a barrier.
In the following section we turn our attention to the superconducting case.
In section IV we compute the Josephson current in the presence of the
modified boundary conditions. Our conclusions are finally stated in section
V. To simplify the notation in the following we will drop the explicit $\mu $
dependence of $r$, and only reinstate it whenever necessary.

\section{Normal conductor with a barrier}

The case of two normal conductors (N-I-N) separated by a barrier was analyzed
by Laikhtman and Luryi \cite{laikhtman94},\cite{note1}. We present here a
derivation of the effective boundary conditions which differs from that
presented in \cite{laikhtman94} and is applicable also to the more general
case when one or both conductors are in the superconducting state (S-I-N and
S-I-S). In the case of normal conductors, eq.(\ref{1}) implies the following
boundary condition

\begin{equation}
f_2 (\mu ) = T f_1 (\mu ) + R f_2 (-\mu )
\label{10}
\end{equation}

where $f_{1,2}= (f_0 + f_z)_{1,2}$ is  the usual  distribution function, as
it is clear by recalling the definitions of the functions $f_0$ and $f_z$  in
eq.(\ref{5}) (see also \cite{volkov96}). Note also that, in the normal case,
eq.(\ref{10}) can be easily derived by means of simple counting arguments. By
considering the symmetric ($s$) and antisymmetric ($a$) parts of $f$, one
rewrites the boundary condition (\ref{10}) in the form

\begin{equation}
-r a(0) =\left[ s \right] \equiv s_2 (0) - s_1 (0),
\label{11}
\end{equation}

where we have made use of the continuity of the antisymmetric part $a$ at the
boundary. 

In order to find a spatial dependence of the distribution function, one has
to solve the Boltzmann kinetic equation,  in the relaxation time
approximation in terms of $s$ and $a$ reads 

\begin{equation}
s' =-a/\mu\equiv - b
\label{12}
\end{equation}
\begin{equation}
\mu^2 b' =<s>-s
\label{13}
\end{equation}

where to make our notation more compact $s'\equiv l\partial_z s$ and, we
assume for simplicity that the mean free path, $l$, is the same on both sides
of the barrier. Eliminating $s$ from  eqs.(\ref{12}-\ref{13}), we can write
the equation for $b$ as

\begin{equation}
\mu^2 b'' -b =-<b> +B_0 \delta (z/l)
\label{14}
\end{equation}

which is valid for all z. The function $B_0 (\mu )\equiv B_0$ determines the
jump in the derivative $b'$ at $z=0$

\begin{equation}
B_0 = \mu^2 \left[ b' \right] = \left[ <s>-s\right]=
r\mu b(0,\mu )-<r\mu b(0,\mu )>
\label{15}
\end{equation}

where the last equality follows from eq.(\ref{11}). We single out the
asymptotic part of $b(z)$ and $s(z)$

\begin{equation}
b(z)=b_{\infty} +\delta b(z),~~~ s(z)=s_{\infty} +\delta s(z)
\label{16}
\end{equation}

in such a way that the functions $b_{\infty}$ and $s_{\infty}$ do not depend
on $\mu$ and are connected to each other via the relation (see eq.(\ref{12}))

\begin{equation}
b_{\infty} = -s'(\pm \infty).
\label{17}
\end{equation}

It follows from eq.(\ref{13}) that the average of $<\mu^2 b>$ does not depend
on the coordinate (this amounts to the conservation of the current) and
therefore can be evaluated from the asymptotic part

\begin{equation}
<\mu^2 b> ={1\over 3} b_{\infty} =<\mu^2 b(0, \mu )>.
\label{18}
\end{equation}

As a result eq.(\ref{14}) can now be written for the function $\delta b$
decaying at $|z|\rightarrow\infty$

\begin{equation}
\mu^2 \delta b'' -\delta b =-<\delta b> +B_0 \delta (z/l).
\label{19}
\end{equation}

By performing a Fourier transform, we obtain for the Fourier component
$\delta b_q$

\begin{equation}
\delta b_q = m_q \left( <\delta b_q > - B_0 \right)
\label{20}
\end{equation}

and hence for the average

\begin{equation}
<\delta b_q > = {{q^2}\over {1-<m_q>}}<m_q\mu^2 B_0 >,
\label{21}
\end{equation}

where $m_q=1/(1+\mu^2 q^2)$, $<m_q >=arctan(q)/q$. By performing the inverse
Fourier transform, we find the magnitude of $b(z, \mu )$ at $z=0$

$$
b(0,\mu )=b_{\infty} +\int_{-\infty}^{\infty}~{{dq}\over {2\pi}} m_q
$$

\begin{equation}
\left( {{q^2}\over {1-<m_q>}}<m_q r(\mu) \mu^3 b (0,\mu )>
-r(\mu)\mu b(0,\mu )\right).
\label{22}
\end{equation}

Eq.(\ref{22}) is an integral equation for $b(0,\mu )$. It can be rewritten in
the form

\begin{equation}
\tilde b (\mu ) (1+r(\mu)/2)= 1+\int_0^1~d\mu_1 r(\mu_1) \mu_1 
{\cal K}(\mu, \mu_1) \tilde b (\mu_1)
\label{23}
\end{equation}

where

$$
\tilde b (\mu) = b(0,\mu)/b_{\infty}
$$

$$
{\cal K}(\mu, \mu_1) = \int_{-\infty}^{\infty}~{{dq}\over {2\pi}} m_q (\mu)
m_q (\mu_1) {{q^2\mu^2_1}\over{1-<m_q>}}.
$$

By a little manipulation of the boundary condition (\ref{11}) we get

\begin{equation}
-r\mu b(0,\mu) =\left[ s_{\infty} \right] +\left[ \delta s \right]
\label{24}
\end{equation}

where $\left[ s_{\infty} \right] \equiv s(\infty) - s(-\infty)$.

Here the jump $\left[ \delta s \right] \equiv \delta s_2 (0) -\delta s_1 (0)$
and taking into account the continuity of $\delta b$ at $z=0$, we obtain from
(\ref{12}) 

\begin{equation} \left[ \delta s \right] = \int_{-\infty}^{\infty}~dz \delta
b(z)=\delta b_{q_0}
\label{25}
 \end{equation}

where $q_0 \equiv 0$. By using eq.(\ref{20}) to express $\delta b_{q_0}$ and
taking into account both eqs.(\ref{24}) and (\ref{25}), after angle
averaging, we arrive at the desired effective  boundary condition

\begin{equation}
-3<r(\mu ) \mu^3 \tilde b (\mu ) > b_{\infty} = \left[ s_{\infty} \right].
\label{26}
\end{equation}

Once the solution of the integral equation $\tilde b(\mu )$ is known,
eq.(\ref{26}) provides a relation between $b_{\infty}$ and the jump of the
symmetric part $\left[ s_{\infty} \right]$. Such a relation is useful in
evaluating the current density $j$, which can be expressed in terms of
$b_{\infty}$ as 

\begin{equation}
j={{\sigma}\over {2el}} \int_{-\infty}^{\infty}~d\epsilon ~ b_{\infty}
\label{27}
\end{equation}

where $\sigma =2e^2 N_0 v_F l/3$ is the conductivity, $N_0$ being the single
particle density of states per spin. If the symmetric part of the
distribution function is in equilibrium on both sides of the barrier, then 

\begin{equation}
\left[ s_{\infty} \right] = th\left[ (\epsilon +eV)\beta \right]-
th\left[ (\epsilon -eV)\beta \right]
\label{28}
\end{equation}

where $2V$ is the voltage drop across the barrier, $\beta =1/2T$. As a
result, the barrier resistance per unit area becomes

\begin{equation}
R_b = (2V/j)={{3l}\over {\sigma}} <r(\mu ) \mu^3 \tilde b (\mu )>.
\label{29}
\end{equation}

This formula gives a relationship between the barrier resistance $R_b$ and
the reflection and transmission coefficients. The function $\tilde b (\mu )$
must be found from eq.(\ref{23}). In the 3-dimensional case eq.(\ref{29})
reads

\begin{equation}
R_b = R_{b0} <r(\mu ) \mu^3 \tilde b (\mu )>
\label{30}
\end{equation}

where

$$
R_{b0} ={{9\pi^2 \hbar }\over {k_F^2 e^2}}.
$$

The integral equation (\ref{23}) must be solved numerically, but before
describing the numerical solution, it instructive to consider few limiting
cases, where an  analytical approach is possible.

\subsection{Weak barrier}

This means that $r \ll 1$ (this condition should be fulfilled for angles not
too close to $\pi /2$). Then, it is seen from eq.(\ref{23}) that 

\begin{equation}
\tilde b (\mu ) \approx 1
\label{31}
\end{equation}

and 

\begin{equation}
R_{bw} =R_{b0} <r \mu^3 >.
\label{32}
\end{equation}

For example, in the case of a thin barrier, modelled by a delta-like
potential at $z=0$, we have that

\begin{equation}
r (\mu ) = s^2 / \mu ^2
\label{33}
\end{equation}

and we obtain

\begin{equation}
R_{bw}=R_{b0} s^2/2
\label{34}
\end{equation}

where $s=\sqrt{2} U_b w / (v_F \hbar )$; $U_b$, $w$, and $v_F$ being  the
barrier height, barrier width, and Fermi velocity, respectively.

\subsection{Strong barrier}

This means that $r \ll 1$. To lowest order the solution to eq.(\ref{23}) is
obtained as 

\begin{equation}
\tilde b (\mu ) = C / (r  \mu )
\label{35}
\end{equation}

where $C$ is a $\mu$-independent constant which can be determined by
exploiting the fact that the average $<\mu^2 b( 0,\mu)>$ does not depend on
$z$. In fact, as follows from eq.(\ref{18})

$$
C=\left( 3 <\mu / r > \right)^{-1}
$$

which lead to a barrier resistance equal to

\begin{equation}
R_{bs} =R_{b0} \left( 9 <\mu / r  > \right)^{-1}.
\label{36}
\end{equation}

The above results coincides with that obtained by Ref.\cite{kuprianov88}, as
can be noted from eq.(\ref{8}). For a thick barrier, we get

\begin{equation}
R_{bs} = R_{b0} {4\over 9} s^2.
\label{37}
\end{equation}

We note that if we were to use, wrongly, eq.(\ref{36}) in the case of a weak
barrier, we would obtain the expression (\ref{37}) instead of (\ref{34}). The
ratio of these two results, $9/8$, is close to $1$. However, this ratio does
not contain any small parameter. This conclusion is in agreement with that of
Ref.\cite{laikhtman94}.

One can obtain a correction $\tilde b_1 (\mu )$ to the solution (\ref{35}).
To see this, we seek a solution in the form

\begin{equation}
\tilde b_1 (\mu ) =\chi (\mu ) / (r \mu ).
\label{38}
\end{equation}

Substituting (\ref{38}) into eq.(\ref{23}), we obtain an integral equation for
$\chi (\mu )$

\begin{equation}
{{1}\over {3\mu r  <\mu /r>}} -1=-{{\chi (\mu )}\over {2\mu }}+
\int_0^1~d\mu_1 {\cal K}(\mu , \mu_1 ) \chi (\mu_1 ).
\label{39}
\end{equation}

It follows from eq.(\ref{39}) that the function $\chi$ is of the order $1$,
i.e. it is small compared to $C$ in eq.(\ref{35}). Therefore, in order to
obtain the effective boundary condition connecting $b_{\infty}$ and $\left[
s_{\infty}\right]$ in the general case, we must solve the integral equation
(\ref{23}).

\subsection{Numerical results}

In the appendix we discuss how the integral equation is solved and  here we
merely illustrate the numerical results. Figure 2 shows the behaviour of the
function ${\tilde b}(\mu )$ for different values of the parameter $s$. We see
that at small and large $s$ the numerics yield the expected behaviour
discussed above. It is perhaps worth noting that for a planar barrier the
Landauer formula yields

\begin{equation}
R_{barrier} = {{h}\over{2e^2}} \left( {{2\pi}\over{{k_F}^2}} \right)
<{{\mu T(\mu )}\over{R(\mu )}}>^{-1}.
\end{equation}

which agrees with eq.(\ref{36}) and is smaller than eq.(\ref{32}) by a factor
of $8/9$. Figure 3 shows the behaviour of the barrier resistance as a
function of $s$ normalized to the $s = 0$ resistance.

In the next section, we now turn to the analysis of the more general case of
a S-I-N or S-I-S interface.

\begin{figure}
{\psfig{figure=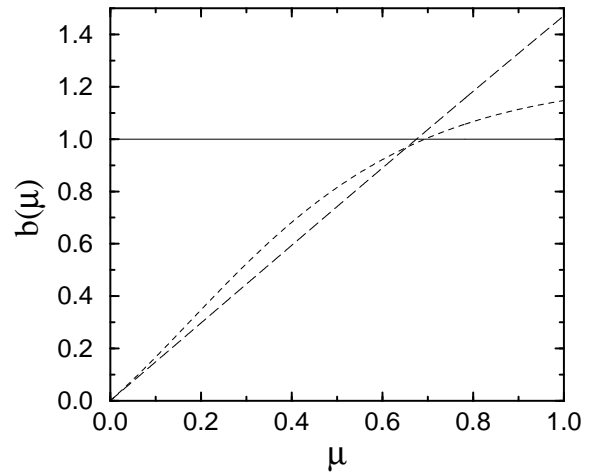,width=8cm}}
\caption{ The solution of the integral equation, ${\tilde b}(\mu )$ is
plotted as function of $\mu $ for different values of $s$:
$s=0$ solid line, $s=1$ dashed line, $s=10$ long-dashed line.
The mesh size in $\mu$ space is 1024 points.}
\label{f2}
\end{figure}

\begin{figure}
{\psfig{figure=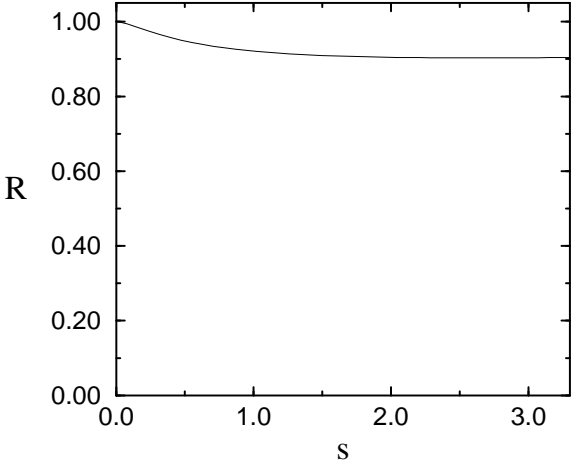,width=8cm}}
\caption{ Barrier resistance $R_b$ versus $s$.}
\label{f3}
\end{figure}

\section{General case: S-I-N and S-I-S}

In this section we analyze boundary conditions for the quasiclassical matrix
Green's function ${\check g}$ in the dirty limit. We will start with the
boundary conditions for ${\check g}$ obtained by Zaitsev in the general case
\cite{zaitsev84} and, as in the preceding section, find a relation connecting
the asymptotic values of ${\check g}$ at $|z|\gg l$ on the two sides of the
barrier. In contrast to the previous case, the presence of anomalous
components in the matrix Green's function leads to a nonlinear equation
governing ${\check g}$. In order to simplify the problem, we note that in the
case of a weak barrier (i.e. $R\rightarrow 0$, $T\rightarrow 1$) the boundary
condition for the symmetric parts ${\check s}_{1,2}$  reduces to a continuity
condition at the interface. It is then physically relevant to confine our
analysis to the most interesting case of a strong barrier and obtain a
relation between ${\check a}$ and ${\check s}$ using the expansion in the
small parameter $r^{-1}$.

By assuming that $r\gg 1$, eq.(\ref{1}) can be cast in the form

\begin{equation}
{\check b}(0,\mu )\approx {1\over { 2r \mu }} 
\left[{\check s}_2 , {\check s}_1 \right] 
\left( 1 -{1\over {2r}} \left( {\check s}_2 - {\check s}_1 \right)^2\right)
\label{40}
\end{equation}

where as in the previous section we introduced the function ${\check b}
={\check a}/\mu$. Eq.(\ref{40}) is valid up to terms of the order $r^{-2}$.
From the equations for ${\check s}$ and ${\check b}$ (see below), one can
obtain the relation corresponding to eq.(\ref{18}), with $b$ replaced by the
supermatrix ${\check b}$. We then proceed as before by writing ${\check s}$
and ${\check b}$ as the sum of a fast decaying and an asymptotic part

\begin{equation}
{\check s}={\check s}_{\infty}+\delta {\check s},~~~
{\check b}={\check b}_{\infty}+\delta {\check b}
\label{41}
\end{equation}

where we assume $\delta {\check s} \ll {\check s}_{\infty}$. By multiplying
eq.(\ref{40}) by $\mu^2$ and using the representation (\ref{41}), we perform
the angle average of eq.(\ref{40}) to obtain

$$
{1\over 3} {\check b}_{\infty} =
\left[ {\check s}_{2\infty} , {\check s}_{1\infty} \right] 
\left( <{{\mu}\over {2r}}>
 -<{{\mu }\over {4r^2}} > \left( {\check s}_{2\infty}
 - {\check s}_{1\infty} \right)^2\right)
$$

\begin{equation}
+<{{\mu }\over {2r}} 
\left( \left[ {\check s}_{2\infty}, \delta {\check s}_{1} \right] +
\left[ \delta {\check s}_{2}, {\check s}_{1\infty} \right] \right) >.
\label{42}
\end{equation}

Here $ \delta {\check s}_{2,1} = \delta {\check s}_{2,1} (0^{\pm})$. The
problem is then reduced to the calculation of the functions $\delta {\check
s}_{2,1}$. We start by writing the equation for $ {\check g}$ in the space
interval $0<|z|\ll \xi_{N,S}$. It is then sufficient to retain only the
gradient term and the  collision integral in the self-consistent Born
approximation for the impurity scattering. As a result the equation for
${\check g}$ reads\cite{larkin86}

\begin{equation}
2\mu {\check g}'={\check g}<{\check g}>-<{\check g}>{\check g}.
\label{43}
\end{equation}
 
By rewriting eq.(\ref{43}) in terms of ${\check b}$ and ${\check s}$ we get

\begin{equation}
2\mu^2 {\check b}'={\check s}<{\check s}>-<{\check s}>{\check s}
\label{44}
\end{equation}

\begin{equation}
 2 {\check s}'={\check b}<{\check s}>-<{\check s}>{\check b}
\label{45}
\end{equation}

together with the conditions deriving from the normalization condition
${\check g}{\check g}={\check 1}$,

\begin{equation}
{\check s}{\check s}=1,~~~~{\check s}{\check b}+{\check b}{\check s}=0.
\label{46}
\end{equation}

Using the expansion (\ref{41}) we obtain the equations for the deviations
$\delta {\check b}$ and $\delta {\check s}$ in the form

\begin{equation}
\mu^2 \delta {\check b}'=-{\check s}_{\infty} \left(
\delta {\check s}-<\delta {\check s}>\right)
\label{47}
\end{equation}

\begin{equation}
  \delta {\check s}'=- {\check s}_{\infty} \delta {\check b}~~~
{\check s}_{\infty}'=-{\check s}_{\infty}{\check b}_{\infty}.
\label{48}
\end{equation}

where we have used the relations

\begin{equation} \delta {\check b}{\check s}_{\infty}+{\check
s}_{\infty}\delta {\check b}=0,~~~ {\check s}_{\infty}\delta {\check s}
+\delta {\check s} {\check s}_{\infty} =0
 \label{49}
 \end{equation}

\begin{equation}
{\check s}_{\infty}{\check s}_{\infty}=1
\label{50}
\end{equation}

which follow from (\ref{46}). From eqs.(\ref{47}-\ref{48}) we finally get the
equation for $\delta {\check b}$

\begin{equation}
\mu^2 \delta {\check b}'' -\delta {\check b}=-<\delta {\check b} > + 
{\check B}_0 \delta (z/l).
\label{51}
\end{equation}

One can easily check that the matrix ${\check B}_0$ is connected to the
Fourier component $\delta {\check b}_{q_0}$ (where $q_0\equiv 0$) by

\begin{equation}
{\check B}_0 =<\delta {\check b}_{q_0}>-\delta {\check b}_{q_0}.
\label{52}
\end{equation}

From eq.(\ref{51}) we find the Fourier components

\begin{equation}
\delta {\check b}_q =m_q \left( {{q^2}\over {1-<m_q>}}
<m_q\mu^2{\check B}_0> -{\check B}_0 \right)
\label{53}
\end{equation}

and the value of $\delta {\check b} (0,\mu )$ at $z=0$ reads

$$
\delta {\check b}(0,\mu )\equiv  {\check b}(0,\mu ) -{\check b}_{\infty} =
$$

\begin{equation}
-\int^{\infty}_{-\infty}~{{dq}\over {2\pi}}~~m_q
\left( {{q^2}\over {1-<m_q>}}
<m_q\mu^2\delta {\check b}_{q_0}> -\delta {\check b}_{q_0} \right).
\label{54}
\end{equation}

To close the above equation, we need to connect ${\check b}(0,\mu )$ and
${\check b}_{\infty}$. To lowest order in $r^{-1}$, the boundary condition
(\ref{42}) yields

$$
{\check b}(0,\mu )\approx (2r\mu)^{-1}
\left[{\check s}_{2\infty}, {\check s}_{1\infty} \right],
$$

\begin{equation}
{\check b}_{\infty} \approx 3<\mu/2r>
\left[{\check s}_{2\infty}, {\check s}_{1\infty} \right].
\label{55}
\end{equation}

which when substituted into (\ref{54}) yields the equation

$$
{\check b}_{\infty}\left( {{1}\over {3\mu r  <\mu /r>}} -1\right) =
$$

\begin{equation}
-\int^{\infty}_{-\infty}~{{dq}\over {2\pi}}~~m_q
\left( {{q^2}\over {1-<m_q>}}
<m_q\mu^2\delta {\check b}_{q_0}> -\delta {\check b}_{q_0} \right).
\label{56}
\end{equation}

If we seek the solution in the form 

\begin{equation}
\delta {\check b}_{q_0} =-\chi (\mu ) {\check b}_{\infty}
\label{57}
\end{equation}

then the function $\chi$ satisfies the integral equation (\ref{39}) and the
Fourier component $\delta {\check b}_{q_0}$ is related to $\delta {\check
s}_{1,2}$ by integrating eq.(\ref{48}) from $-\infty$ to $0$ and from $0$ to
$\infty$, to yield

\begin{equation}
\delta {\check s}_2 = {\check s}_{2\infty}\delta {\check b}_{q_0}/2,~~~
\delta {\check s}_1 = -{\check s}_{1\infty}\delta {\check b}_{q_0}/2.
\label{58}
\end{equation}

Substituting eqs.(\ref{58}), (\ref{57}), and (\ref{55}) into eq.(\ref{42}) we
finally obtain

$$
{{ {\check b}_{\infty}}\over 3} =
\left[{\check s}_{2\infty} , {\check s}_{1\infty} \right] 
\left( <{{\mu}\over {2r }}>
 -<{{\mu }\over {4r^2}} >\left( {\check s}_{2\infty}
 - {\check s}_{1\infty} \right)^2\right)
$$

\begin{equation}
+3 <{{\mu\chi }\over {2r}}><{{\mu }\over {2r}} >
\left[{\check s}_{1\infty} ,
 {\check s}_{2\infty}{\check s}_{1\infty}{\check s}_{2\infty} \right]  .
\label{59}
\end{equation}

The above equation is the effective boundary condition for the matrix
${\check g}$ in the dirty limit. The first term in (\ref{59}) coincides with
the boundary condition obtained in Ref.\cite{kuprianov88}. 

\section{The Josephson effect}

As a simple application of the above boundary condition, we now derive an
expression for the Josephson current. To this end we rewrite eq.(\ref{59}) in
the following way

\begin{equation}
{1\over 3} {\check b} = A [{\check g}_2,{\check g}_1] +
B  [{\check g}_2,{\check g}_1]  \lbrace{\check g}_2,{\check g}_1\rbrace
\label{60}
\end{equation}

where we have identified the symmetric part ${\check s}$ with the Green's
function ${\check g}$ and dropped the $\infty$ suffix and the curly brackets
indicate the anticommutator

$$
 \lbrace{\check g}_2,{\check g}_1\rbrace=
 {\check g}_2 {\check g}_1+
 {\check g}_2 {\check g}_1.
$$

The constants $A$ and $B$ can be read off from eq.(\ref{59})

$$
A=<\mu /r>-2<\mu/4r^2>
$$
and

$$
B=<\mu /4r^2>-3<\mu\chi/2r><\mu/2r>.
$$

In deriving eq.(\ref{60}) we have made use of the normalization condition
${\check g}{\check g}={\check 1}$. The current through the junction is
determined by the formula

\begin{equation}
I=-{{\sigma }\over {16el}}\int_{-\infty}^{\infty} ~d\epsilon 
Tr( {\hat \sigma}_z {\hat b})
\label{61}
\end{equation}

where ${\hat b}$ is the appropriate Keldysh component. In the absence of a
voltage across the junction, the Keldysh component of the supermatrix
${\check g}$ reduces to

\begin{equation}
{\hat g}=f_0 ({\hat g}^R-{\hat g}^A)
\label{62}
\end{equation}

where $f_0=2tanh(\epsilon /2T)$ is the equilibrium distribution function. The
Keldysh component of the product of the commutator and the anticommutator
reads

\begin{equation}
( [{\check g}_2,{\check g}_1] \lbrace{\check g}_2,{\check g}_1\rbrace)_k=
 [{\hat g}_2^R,{\hat g}_1^R] \lbrace{\check g}_2,{\check g}_1\rbrace_k+
 [{\check g}_2,{\check g}_1]_k \lbrace{\hat g}_2^A,{\hat g}_1^A\rbrace
\label{62b}
\end{equation}

with the Keldysh component of the commutator

\begin{equation}
[{\check g}_2,{\check g}_1]_k =f_0 (
[{\hat g}_2^R,{\hat g}_1^R] -[{\hat g}_2^A,{\hat g}_1^A]  )
\label{63}
\end{equation} 

and of the anticommutator

\begin{equation}
\lbrace {\check g}_2,{\check g}_1\rbrace_k =f_0 (
\lbrace {\hat g}_2^R,{\hat g}_1^R\rbrace -
\lbrace {\hat g}_2^A,{\hat g}_1^A\rbrace  ).
\label{64}
\end{equation} 

To calculate the current we represent the matrices in the Nambu space as

$$
{\hat g}^{R(A)}_{1,2}=G^{R(A)} {\hat \sigma}_z
+iF^{R(A)}(cos(\phi_{1,2}) {\hat \sigma}_y+
sin(\phi_{1,2}) {\hat \sigma}_x )
$$

where $\phi_{1,2}$ are the phases of the superconducting order parameter on
the two sides of the junction. The current in eq.(\ref{61}) can be written
then as the sum of three terms

\begin{equation}
I_A=-iI_{0,A} sin(\phi_1 -\phi_2 )
\int_{-\infty}^{\infty} d\epsilon
f_0(F^R_1F^R_2-F^A_1F^A_2 ),
\label{65}
\end{equation}

$$
I_B^{(1)}=-iI_{0,B}2sin(\phi_1 -\phi_2 )
$$

\begin{equation}
\int_{-\infty}^{\infty} d\epsilon
f_0(F^R_1F^R_2G^R_1G^R_2-F^A_1F^A_2G^A_1G^A_2 ),
\label{66}
\end{equation}

and 

$$
I_B^{(2)}=-iI_{0,B}sin(2(\phi_1 -\phi_2 ))
$$

\begin{equation}
\int_{-\infty}^{\infty} d\epsilon
f_0((F^R_1F^R_2)^2-(F^A_1F^A_2)^2 ).
\label{67}
\end{equation}

In the above formulae $I_{0,A}=(eN_0v_F/2)A$, $I_{0,B}=(eN_0v_F/2)B$, $e$ is
the electron charge, $N_0$ the single particle density of states per spin,
and $v_F$ the Fermi velocity. By using the expression for $G^{R(A)}$ and
$F^{R(A)}$ at equilibrium 

$$
G^{R(A)}={{\epsilon}\over {\sqrt{(\epsilon_{\pm})^2-\Delta^2}}},
F^{R(A)}={{\Delta}\over {\sqrt{(\epsilon_{\pm})^2-\Delta^2}}},
$$

where $\epsilon_{\pm}\equiv \epsilon \pm i0^+$, and assuming that the gap
$\Delta$ is equal on both sides of the junction, we obtain, at $T=0$, the
following result for the current

\begin{equation}
I=eN_0v_F\Delta\pi\left[ (2A+B)sin(\phi )-Bsin(2\phi )\right]
\label{68}
\end{equation}

where $\phi=\phi_2-\phi_1$. Note that by confining ourselves to the lowest
order in $r^{-1}$, we would obtain for the Josephson current the standard
result of tunneling theory

\begin{equation}
I=(e^2N_0v_F<\mu /r>){{\pi\Delta}\over {2e}}sin(\phi )
=R_{bs}^{-1}{{\pi\Delta}\over {2e}}sin(\phi )
\label{69}
\end{equation}

with $R_{bs}$ given by eq.(\ref{37}). Allowance for higher order terms in the
barrier transparency leads then to higher harmonics in the current phase
relation. This result has a simple physical interpretation. We know that in
the case of a superconductor - normal metal - superconductor structure, the
Josephson effect manifests itself with a triangular shape of the current -
phase relation. For this case  the Fourier decomposition has an infinite
number of harmonics. Hence it is clear that higher order terms in the
$r^{-1}$ expansion must possess harmonics of higher order. It may be useful
at this point to notice that very recently Josephson current measurements
have been carried out in structures in which it is possible to control
experimentally the barrier transparency. The experimental results show indeed
that by tuning the barrier strength it is possible to observe higher
harmonics \cite{lt21}.

\section{Conclusions}

We have analyzed the boundary condition for the quasiclassical matrix Green's
function ${\check g}$ at the S-N or S-S' boundaries in the dirty limit.
Effective boundary conditions have been derived with the aid of an expansion
in the barrier transmittance. The first term of the expansion reproduces the
results previously derived by Kupriyanov and Lukichev \cite{kuprianov88}. In
the normal case, the boundary conditions for a barrier of arbitrary
transparency may be obtained from the solution of an integral equation. In
the superconducting case, due to the nonlinear nature of the equations, we
have been able to compute the next-to-leading term in the $r^{-1}$ expansion.
In this case the evaluation of this term also entails solving an integral
equation. To illustrate the physical consequences of the modified boundary
condition, we have calculated the Josephson current for a tunnel junction
between two superconductors at equilibrium and shown that higher order
harmonics arise. Given the relevance that the lowest order boundary
expression (cf. eq.(\ref{8})) has played in the study of mesoscopic
normal-superconducting hybrid structures, our result calls for a
reexamination of the various effects occurring in mesoscopic hybrid
structures. In particular, one may envisage a straightforward generalization
of the circuit theory of Nazarov \cite{nazarov94} to allow for the new
boundary condition. This will be the subject of a future investigation.

\acknowledgements
Financial support from the EPSRC is gratefully acknowledged.

\appendix
\section{Solving the integral equation: technical details}

The variable $\mu$ is discretized $\mu_i =i\Delta_\mu $ with $i=1, ...,N$ and
$\Delta_\mu=1/N$. By introducing a vector $b_i = {\tilde b}(\mu_i )$, the
integral equation $(\ref{23})$ acquires the matrix form

\begin{equation}
b_i (1+r_i/2)=1+\sum_{j=1}^N K_{ij}r_j\mu_j b_j
\label{a1}
\end{equation}

where $r_i=r(\mu_i )$ and 

\begin{equation}
K_{ij} =\int_{-\infty}^{\infty}~{{dq}\over {2\pi}} m_q (\mu_i )
m_q (\mu_j ) {{q^2\mu^2_j}\over{1-<m_q >}}.
\label{a2}
\end{equation}

The solution $b_i$ can then be obtained as

\begin{equation}
b_i = \sum_{j=1}^N (A^{-1})_{ij}
\label{a3}
\end{equation}

where the matrix $A_{ij}$ is given by

\begin{equation}
A_{ij}=(1+r_i/2)\delta_{ij}-\Delta_\mu K_{ij}r_j\mu_j .
\label{a4}
\end{equation}

$^{a}$ Permanent address: Dipartimento di Fisica E. Amaldi, Terza
Universit\`a degli Studi di Roma, via della Vasca Navale 84, I-00146 Roma,
Italy.

\end{document}